# Nanoacoustic Defect Manipulation in Solids.

Igor Ostrovskii[1], Nataliya Ostrovskaya[1], Oleg Korotchenkov[2], and James Reidy[1]

[1]Department of Physics and Astronomy, University of Mississippi, University MS 38677.
[2] General Physics Department, Kiev Shevchenko University, Kiev-01022, Ukraine.

**Abstract.**

Within a nanoscale volume, an acoustic wave interacts with radiation defects in ionic solids. The radiation-induced optical absorption in ionic crystals is remarkably removed by a room-temperature ultrasonic treatment of the crystals. It is shown that the effect can be explained by defect migration processes occurring in ultrasonic fields. The inter-ion charge exchange, termed as acoustically-stimulated chemical reaction, is furthermore suggested to occur affecting the defect migration. This new method of a cold annealing of radiation defects in solids can be regarded as nanoacoustic defect manipulation.

## 1. Introduction.

The experimental and theoretical investigations of the properties of materials exposed to ionizing and nuclear radiation remains to be an active area of research and development. Among numerous studies done, one can mention radiation-related changes in optical properties of ionic crystals [1 - 3] and II-VI compounds [4], and radiation-induced generation of point defects in semiconductors [5- 7] which can influence the electrical characteristics of silicon-based detectors [8]. The crystals are widely used as scintillator materials in detector and calorimeter systems for the purposes of high-energy physics and medicine. In these applications, they are exposed to nuclear radiation of different origin, resulting in an increased concentration of point defects and their agglomerates. The defect generation leads to the degradation of material properties and device performance [9].

In ionic crystals, different types of color centers are typically generated by the high-energy radiation. In particular, they cause the deterioration of transmission characteristics of scintillators due to the appearance of absorption bands and the overall decrease in optical transmission. Therefore, the developments of the techniques able to remove the induced absorption are of significant importance. Meanwhile, there are no too many possibilities to do that. The induced absorption is traditionally removed by thermal annealing in the temperatures range from 100 to 900 $^0$C, depending on the crystal used.

An attractive alternative route can offer a defect manipulation approach, sometimes referred to as "defect engineering" [10-13]. For example, this respectively new method was successfully applied to quench an excessive concentration of dopants, mainly oxygen in case of silicon [10].

This work aims to apply the method of nanoacoustic defect manipulation (NDM) by ultrasound in order to remove the absorption in ionic crystals induced by X rays and γ-irradiation. The idea of the defect manipulation approach consists of the ultrasound-mediated activation of the defect diffusion thus altering the concentration and the space location of point defects in a given crystal volume including the near surface region of the crystal.

It is therefore the purpose of this work to achieve cold annealing of radiation defects without the sample heating. This would have particular relevance to detector technologies and performance of calorimetric systems. In many cases, when the device is a rather complicated system consisting of a crystal on a chip, bounding wiring, etc. the high temperature can destroy some elements of the structure. Then the NDM is thought to be applicable as the ultrasound source can easily be integrated into the multicomponent detectors or detector array systems. The NDM can be applied not only to the single crystals damaged by high-energy radiation but also to the real detectors or detector array systems.

**2. Defect Manipulation.**

We use the single crystals of CsI(Tl), CsI(Na), and KBr in the experiments. The samples of KBr and CsI(Na) are from the same vendor (Eastern Europe). The samples having a form of parallelepiped are cut out from a massive piece, the dimensions are about 5 by 5 mm$^2$ for end faces, and a length is about 10 - 12 mm. For the purpose of comparison, we also take optical spectra with the samples of CsI(Tl) from another vendor (Western Europe). Sample CsI(Tl)-2 is made from the original material used in SLAC (Stanford Linear Accelerator) for the crystal detectors of nuclear radiation. In this last case we have respectively large samples (2.54 mm in diameter and length).

The crystals have been exposed to X- or γ- irradiation with the doses which allow to visibly detecting the presence of color centers. Before NDM, the samples are seen good colored, slightly brown in case of CsI and dark blue in case of KBr.

Acoustic waves in the megahertz frequency range are used to cure the samples, which were preliminary irradiated by X rays. All experiments are done at room temperature. Nevertheless, during ultrasonic load, the sample temperature is slightly elevated. Thus, in these studies, the maximum sample overheat caused by ultrasound amounts to ≈30 $^0$C. However, it should be noted that, in this particular case, such a partial conversion of the ultrasonic energy to the heat can be useful effect, activating the point defect migration process. It is also worth noting that the temperatures below ≈550K can be defined as low temperatures for defect annealing.

The experimental setup is shown in Fig. 1. Piezoceramic (PZT) ultrasonic transducers 1 and 3 are bounded to the sample 2. An rf-voltage ($V_{RF}$) in the MHz frequency (f) range is applied to the input transducer 1, which converts this voltage into ultrasonic wave. By measuring an output rf-voltage ($V_{OUT}$), one can find an optimum frequency for maximum acoustic amplitude inside the sample 2.

To detect the changes in the concentration of point defects, optical transmission spectra are measured using a Perkin-Elmer UV/VIS Lambda 18 double beam spectrophotometer. The spectra are taken before radiation treatment, during ultrasound load, and finally after acoustic wave is removed from the samples.

### 3. Experimental results.

The optical transmission spectra taken in CsI(Na) and CsI(Tl) samples are shown in Fig. 2. Plots 1 and 2 display initial transmittance, whereas plots 3 and 4 illustrate the deterioration of the transmittance after X-ray irradiation. Actually as closer to irradiated region the optical spectra are taken as higher is a decrease in optical transmittance. By numerous investigations, it has been found that after irradiation the optical transmittance drops, so that only 1/3 to 1/2 of the initial transmission of light is observed with the irradiation dose used. Also the displacement of UV absorption edge is observed. Applying NDM, we are able to remarkably remove the radiation-induced absorption. The size of the effect depends on the NDM parameters, as will be discussed below. In CsI, the maximum effect is illustrated by plot 5 in Fig. 2. It is seen that about 60-70% of the radiation-induced absorption can be removed by NDM.

To reveal which absorption bands occur in X-ray irradiated CsI(Tl), we measure the optical spectra under slit spectral width of 2.0 nm with a scan speed of 240 nm/min. The results taken from the CsI(Tl)-sample-2 are shown in fig. 3. The pre-irradiation spectrum of CsI(Tl) (plot 1) exhibits two very weak absorption bands peaked at 460 and 840 nm. After X-ray irradiation four absorption bands at 430, 520, 560, and 840 nm are resolved in the spectrum (fig. 3, plot 2). In some measurements, the small peaks at ~ 350 and 390 nm are also detected. These data are in agreement with the previously reported color centers in doped CsI crystals [1, 14, 15]. It is interesting to note, after X-ray irradiation, the initially strongest peak at 460 nm disappears and another peak at 430 nm appears. The same behavior was observed after a small dose of γ-ray irradiation [14, 15]. The CsI(Tl) spectra turn out to be too complex to allow any short quantitative consideration in the frame of this publication.

The quantitative analysis involving the influence of NDM on the concentration of F-centers can much better be done in γ-irradiated KBr crystals (Figs. 3 and 4). The transmission minimum at about 650 nm seen in Fig. 4 corresponds to the absorption band of F-centers in KBr [16]. Applying NDM, the transmission is improved, depending on the amplitude of the NDM (plots 2 and 3 in Fig. 4). It is also seen in Fig. 4 that the NDM effect exhibits a threshold behavior. Thus, at $V_{RF}$ = 17 Volts small changes are only observed in Fig. 4, whereas at $V_{RF}$ = 18 Volts a significant improvement in optical transparency is seen.

### 4. Discussion.

The experimental results presented in Figs. 2- 4 illustrate the annealing effect of ultrasound on the radiation defects in ionic solids. In order to understand the likely origin of the observed decrease in the defect concentration due to NDM, one can analyze the amplitude characteristics of NDM. It is seen in Fig. 4 that the transmission recovery is observed just above $V_{RF} \approx 17$ Volts. At lower ultrasonic amplitudes, no significant change in optical transmission occurs. Therefore, the above-threshold behavior of NDM is evident.

According to thermodynamic consideration, a concentration of electrically charged vacancies and interstitial atoms is a function of external pressure P,

$$n_+(P) = n_-(P) = (N_+N_-)^{1/2} \exp[\frac{S_{F+} + S_{F-}}{2k} - \frac{E_+ + E_-}{2kT} - \frac{P(V_{F+} + V_{F-})}{2kT}], \quad (1)$$

where: $n_{+,-}$ is a concentration of point defects having positive or negative electric charge; $N_{+,-}$ is a concentration of corresponding regular sites in a crystal lattice; $S_{F+,-}$ and $E_{+,-}$ are the entropy and energy of defect formation. In case of NDM, the pressure exerted by ultrasound on a micro volume of a sample is a sine function of time **P ~ Sin(ωt);** but due to exponential character of amplitude dependence given by equation (1), a net result of NDM is not theoretically (thermodynamically) zero. In other words, when the interstitial atom goes to its native site in one half cycle of acoustic wave, it still remains at its regular site in the next opposite half cycle of acoustic wave. This is due to the fact that the ultrasonic energy is much less than the nuclear radiation energy. Therefore, the former is considered to be not large enough to remove the atom from its regular site, whereas it is thought to be remarkably large to move the atom from the interstitial position.

We can then substitute the acoustically exerted pressure, **P = P₀Sin(ωt)**, by its module value, **|P₀Sin(ωt)|**, in Eq. (1). It can then be concluded that UDM can actually decrease a defect concentration due to ultrasonically stimulated diffusion process.

Within the suggested model, the F-center concentration $N_F$ is

$$N_F \sim \exp[\frac{S_{F+} + S_{F-}}{2k} - \frac{E_+ + E_-}{2kT} - \frac{P(V_{F+} + V_{F-})}{2kT}], \quad (2)$$

where the pressure P is proportional to the amplitude of acoustic wave, which in turn is proportional to the rf-voltage exciting ultrasound (P ~ $CV_{RF}$), where C is a certain coefficient depending on transducer and crystal properties. If one writes an effective volume of the F-center (or color center in general) consisting of a single interstitial halogen ion and its vacancy as $2V_C$, the final dependence resulting from Eqs (1) and (2) is simply

$$N_F^{-1} \sim \exp[C\, V_{RF}\, V_C / kT]. \qquad (3)$$

This dependence is not linear suggesting the existence of a strong interaction, which governs ultrasonic action in case of irradiated ionic crystals. By eqn. (3), we estimate the volume $V_C$, in which the interaction of acoustic wave and a single defect takes place. It appears to be the volume $V_C$ is of nanoscale, and ultrasound transfer its energy to a nano-volume scale defect. Thus NDM is a phenomenon of nano-acoustic interaction in irradiated ionic crystals. To our mind one of the physical explanations of this interaction might be a quantum diffusion stimulated by ultrasound. To support this approach, we can refer to the publication [17], which describes the quantum diffusion in a kicked system. In our case, the sample is periodically activated or kicked by ultrasound, which effectively can be represented by a local increase of temperature due to acoustic mechanical pressure during a compression cycle. One can also consider a tunnel effect, which is activated by external acoustic action. For example, the energy level of interstitial halogen ion is elevated in acoustic fields, so that it can experience the tunneling transition to the closest regular site. For this process, an increase in high-frequency phonon density is needed, which can be realized through a nonlinear acousto-phonon interaction.

Another possible mechanism of the observed diffusion likely to account the charge state of the radiation defects. It has previously been shown experimentally [18] that, during ultrasonic treatment of semi-conducting crystals, a transfer of electrical charges between point defects is possible, changing the electronic subsystem of the crystal itself. Further, it has also been shown [19] that an electronic excitation can in turn stimulate atomic migration in semiconductors. These processes can obviously not be excluded when analyzing the defect migration in ionic crystals.

We will discuss the physics behind the NDM in more details elsewhere. The present work aims to experimentally show the existence of nano-acoustic interaction and subsequent occurrence of ultrasonically stimulated defect migration during NDM in ionic crystals with radiation defects. A detail analysis of the physical mechanisms responsible for atomic migration in our experimental samples remains to be done.

**Conclusions.**

1. It is shown for the first time that within a nanoscale volume, an acoustic wave interacts with radiation defects in ionic solids.

2. The radiation-induced optical absorption in ionic crystals can be remarkably removed by ultrasonic treatment of the crystals at room temperature. This new method of a cold annealing of radiation defects in solids can be regarded as nanoacoustic defect manipulation (NDM) by ultrasound.

3. The effect of NDM is likely to be due to defect migration in ultrasonic fields accompanied by the ultrasound-mediated inter-atomic charge exchange. Physically this process is a type of stimulated by ultrasound chemical reaction in irradiated crystal.

**References:**

[1] M. Nikl, "Wide band gap scintillation materials: Progress in the technology and material understanding," *Phys. Stat. Sol.(a)*, vol. 134, pp. 595-620, April 2000.

[2] A. N. Belsky, A. N. Vasil'ev, and V. V. Mikhailin, "Experimental study of the excitation threshold of fast intrinsic luminescence of CsI," *Phys. Rev. B*, vol. 49, pp. 13197-13200, May 1994.

[3] C. Medrano P., S. Ramos B., J. Hernandez A., H. Murietta S., C. Zaldo, and J. Rubio O., "Influence of radiation intensity and lead concentration in the room-temperature coloring of KBr," *Phys. Rev. B*, vol. 32, pp. 6837-6844, Nov. 1985.

[4] N.D. Vakhnyak, S.G. Krylyuk, Yu.V. Kryuchenko, I.M. Kupchak, "Influence of $\gamma$-irradiation on photoluminescence spectra of CdTe:Cl," *Semicond. Phys. Quantum Electronics & Optoelectronics*, vol. 5, pp. 25-30, Jan. 2002.

[5] B.G. Svenson, C. Jagadish, And J.S. Williams, "Generation of point defects in crystalline silicon by MeV heavy ions: Dose rate and temperature dependence," *Phys. Rev. Let.*, vol. 71, pp. 1860-1863, Sept. 1993.

[6] D.N. Seidman, R.S. Averback, P.R. Okamoto, and A.C. Baily, "Amorphization processes in electron and/or ion-irradiated silicon," *Phys. Rev. Let.*, vol. 58, pp. 900-903, March 1987.

[7] C.S. Chen, J.C. Corelli, and G.D. Watkins, "Radiation-Produced Absorption Bands in Silicon: Piezospectroscopic Study of a Group-V Atom-Defect Complex," *Phys. Rev. B*, vol. 5, pp. 510-526, Jan. 1972.

[8] M. Moll, E. Fretwurst, G. Lindström, "Leakage current of hadron irradiated silicon detectors – material dependence," Nuclear Instruments and Methods in Physics Research. 1999, Sec. A, pp. 87-93.

[9] K. Kazui at al. "Study of the radiation hardness of CsI(Tl) crystals for the BELLE detector", *Nuc. Instrum. and Methods in Phys. Res*. A vol. 394, p. 46-56, 1997.

[10] G. Lindström et al. (RD48), "Radiation hard silicon detectors – developments by the RD48 (ROSE) collaboration", presented at the 4th STD Hiroshima Conf., Hiroshima, Japan, 2000; *RD48 3rd Status Rep*. CERN/LHCC 2000-009, Dec.99.

[11] I.V.Ostrovskii, Ju.M.Khalack, A.B.Nadtochii, H.G. Walther, "Defects Modification by Ultrasound in Crystals," *Solid State Phenomena*, vol.67-68, pp.497-502, 1999.

[12] I. Dirnstorfer, W. Burkhardt, B. K. Meyer, S. Ostapenko, and F. Karg "Effect of ultrasound treatment on CuInSe$_2$ solar cells", *Solid State Communications*, vol.116, pp. 87-91, Sept. 2000.

[13] B. Romanjuk, D. Krüger, V. Melnik, V. Popov, Ya. Olikh, V. Soroka, O. Oberemok, "Ultrasound effect on radiation damages in boron implanted silicon," *Semicond. Phys. Quantum Electronics & Optoelectronics*, vol.3, pp. 15-18, Jan. 2000.

[14] M.A.H. Chowdhury et al. "Radiation effects in CsI(Tl) crystals from a controlled grouth process", *Nucl. Instrum. and Meth. Phys. Res. A*, vol.413, 471-4, 1998.

[15] M.A.H. Chowdhury, S. J. Watts, D.C. Imre, A.K. McKemey, and A.G. Holmes-Siedle, *Nucl. Instrum. and Meth. Phys. Res*. A vol. 432, 147, 1999.

[16] W. B. Fowler, *Physics of Color Centers*. New York: Academic Press, 1968.

[17] P. Facchi, S. Pascazio, and A. Scardicchio, Measurement-Induced Quantum Diffusion, *Phys. Rev. Lett.*, vol.83, N1, p. 61, 5 July 1999.

[18] I. V. Ostrovskii, A. Rozko, "Acoustic redistribution of defects in crystals," *Sov. Phys. Solid State*, vol.26, pp.2241-2242, Dec. 1984.

[19] A. P. Zdebskii, N. V. Mironyuk, S. S.Ostapenko, A. U. Savchuk. and M. K. Sheinkman, "Mechanism for ultrasound stimulated variation of photoelectric and luminescent properties in CdS," *Sov. Phys. Semiconductors*, vol. 20, p. 1167, 1986.

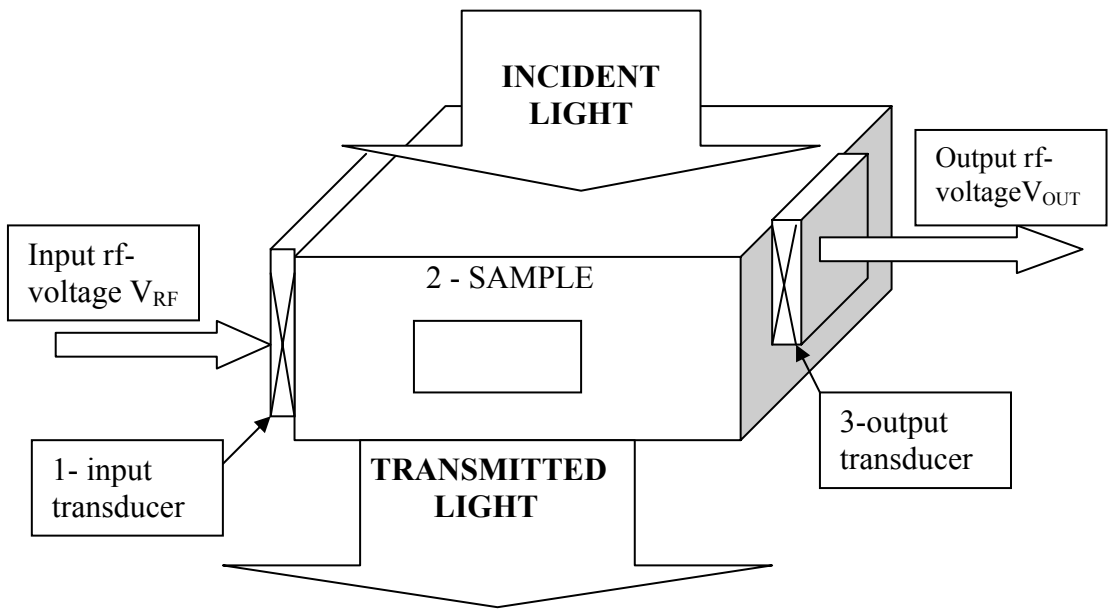

Fig.1. Experimental setup: sketch of NDM and optical measurements: 1- input piezo-ceramic transducer for acoustic wave excitation in a ample 2, 3 – output piezo-ceramic transducer.

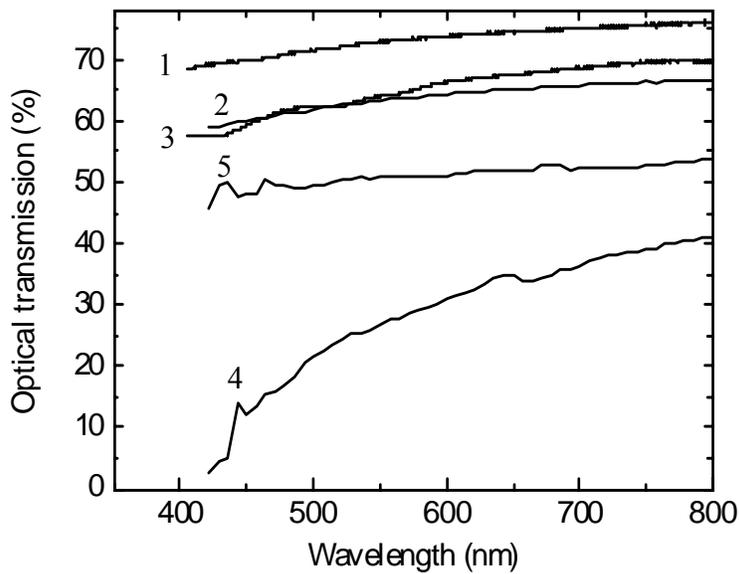

Fig. 2. Influence of X-ray irradiation and NDM on optical transmittance in CsI(Tl) (1, 3) and CsI(Na) (2, 4, 5) crystals. 1, 2 - before X-ray irradiation, 3, 4 - after X-ray irradiation (no NDM), 5 - after NDM of irradiated sample (f = 2.4 MHz, $V_{RF}$=10.5 $V_{eff}$).

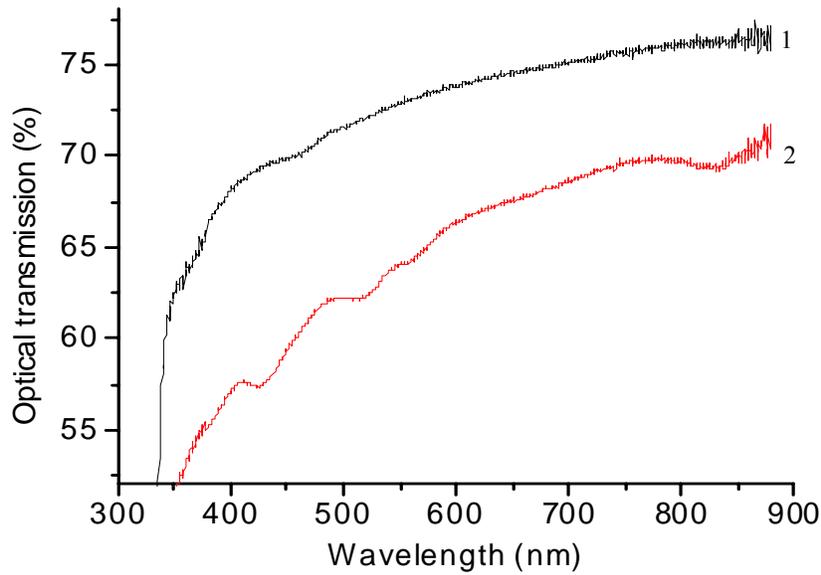

Fig. 3. Influence of X-ray irradiation on optical transmittance in CsI(Tl)-sample-2. Plot 1 – from the area without X-ray irradiation, plot 2- from the area close to damaged region occurring after X-ray irradiation.

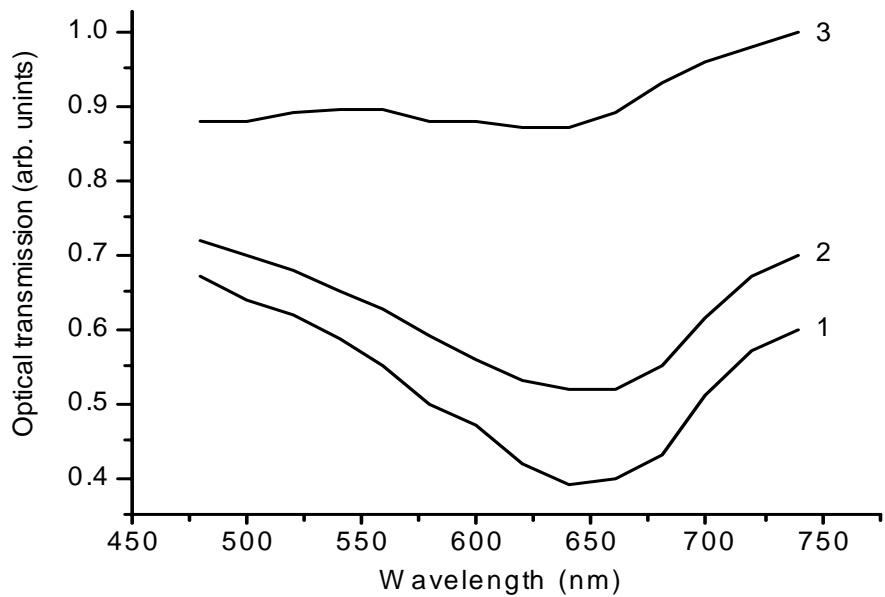

Fig. 4. Radiation-induced transmission spectrum minimum (1) in KBr and its recovery by NDM with $V_{RF}$= 17 (2) and 18 (3) $V_{eff}$, f = 2.16 MHz.